\documentclass[aps,showpacs,showkeys,preprintnumbers,amsmath,amssymb]{revtex4}
\usepackage{dcolumn}
\usepackage[dvips]{epsfig}
\def \be {\begin{equation}}
\def \ee {\end{equation}}
\def \bea {\begin{eqnarray}}
\def \eea {\end{eqnarray}}

\begin{document}
\baselineskip=0.8 cm
\title{\bf  Holographic entanglement entropy in metal/superconductor phase transition with Born-Infeld electrodynamics}

\author{Weiping Yao, Jiliang Jing\footnote{jljing@hunnu.edu.cn}}
%\email{jljing@hunnu.edu.cn}

\affiliation{Department of Physics, and Key Laboratory of Low Dimensional Quantum Structures and
Quantum Control of Ministry of Education, Hunan Normal University,
Changsha, Hunan 410081, P. R. China}

\begin{abstract}
\baselineskip=0.6 cm
\begin{center}
{\bf Abstract}
\end{center}
We investigate the holographic entanglement entropy in the metal/superconductor phase transition for the Born-Infeld electrodynamics with full backreaction and note that the entropy is a good probe to study the properties of the phase transition. For the operator $<\mathcal{O}_{-}>$, we find that the entanglement entropy decreases (or increases) with the increase of the Born-Infeld parameter $b$ in the metal (or superconducting) phase.  For the operator $<\mathcal{O}_{+}>$, we observe that, with the increase of the Born-Infeld parameter, the entanglement entropy in the metal phase decreases monotonously but the entropy in the superconducting phase first increases and forms a peak at some threshold $b_{T}$, then decreases continuously. Moreover, the value of $b_{T}$ becomes smaller as the width
of the subsystem $A$ decreases.
\end{abstract}


\maketitle

\newpage

\section{Introduction}
As the holographic principle provides a new insight into the
investigation of strongly interacting condensed matter systems
\cite{G.'t Hooft1993,L.Susskind1994}, there are a lot of works
applying the anti-de Sitter/conformal field theory (AdS/CFT) duality
\cite{J.Maldacena1998, EWritten1998, S.S.Gubser1998} to condensed
matter physics and in particular to superconductivity
\cite{Ge-Wang-Wu,gregorysoda,Nakano-Wen,Amado,
Koutsoumbas,Umeh,Sonner,Franco,Herzog-2010,HorowitzPRD78,
 Konoplya,Siopsis,maeda,CaiNie,panwang, Oriol Dom¨¨nech2010, jingpanl}.
It states that the bulk AdS black hole becomes unstable and scalar
hair condenses as one tunes the temperature for black hole.
 As a matter of fact, in order to understand the
influences of the $1/N$ or $1/\lambda$ ($\lambda$ is the 't Hooft
coupling) corrections on the holographic superconductors, the higher derivative corrections to the gauge field
should be taken into consideration. For the high order correction
related to the gauge field, one of the important nonlinear
electromagnetic theories is Born-Infeld electrodynamics
\cite{M.Born1934, G.W.Gibbons1995, B.Hoffmann1935, W.Heisenberg1936,
Oliveira1994, O.Miskovic2011}.  As is well known, the Born-Infeld
electrodynamics, which was proposed in 1934 by Born and Infeld to avoid the infinite
self-energies for charged point particles arising in Maxwell theory
\cite{M.Born1934}, displays good physical properties including the
absence of shock waves and birefringence. It was also found that the
Born-Infeld electrodynamics is the only possible nonlinear version
of electrodynamics that is invariant under electromagnetic duality
transformation \cite{G.W.Gibbons1995}. Jing and Chen observed that
the Born-Infeld coupling parameter make it harder for the scalar
condensation to form \cite{Jing-Chen}. Then,
the analytic study of properties of holographic superconductors in Born-Infeld electrodynamics
was presented in Ref. \cite{Sunandan Gangopadhyay2012}. In this paper, We would like to
investigate the phase transition in the Born-Infeld electrodynamics with
full backreaction of the matter fields electrodynamics on the AdS black hole geometry.

Recently, according to the AdS/CFT duality, Ryu and Takayanagi
\cite{Ryu:2006bv, Ryu:2006ef} have presented a proposal to compute
the entanglement entropy of conformal field theories (CFTs) from the
minimal area surface in gravity side. Since this proposal provides a
simple and elegant way to calculate the entanglement entropy of a
strongly coupled system which has a gravity dual, the holographic
entanglement entropy is widely used to study various properties of
holographic superconductors at low temperatures
\cite{Nishioka:2009un,Albash:2011nq,Myers:2012ed,deBoer:2011wk,Hung:2011xb,Nishioka:2006gr,
Klebanov:2007ws,Pakman:2008ui, E. Arias2013}. The entanglement entropy in the
metal/superconductor system  was studied in \cite{T-6}. It was shown
that the entanglement entropy in superconducting phase is always
less than the one in the metal case and the entropy as a function of
temperature is found to have a discontinuous slop at the transition
temperature $T_{c}$ in the case of the second order phase
transition. Ref. \cite{Ogawa:2011fw} considered the case with higher
derivative corrections and studied the holographic entanglement
entropy in Gauss-Bonnet gravity. Ref. \cite{Xi Dong2013} studied the
holographic entanglement entropy for general higher derivative
gravity and proposed a general formula for calculating the
entanglement entropy in theories dual to higher derivative gravity.
Then, the holographic entanglement entropy in the
insulator/superconductor model was studied in \cite{R. G. Cai2012,
Li-Fang Li2013, Cai1203, Cai1209} and it turned out that the
entanglement entropy is a good probe to investigate the holographic
phase transition. Furthermore,  Kuang \emph{et al.} examined the
properties of the entanglement entropy in the four-dimensional AdS
black hole and found that the entanglement entropy can be considered
as a probe of the proximity effect of a superconducting system by
using the gauge/gravity duality in a fully backreacted gravity
system\cite{Kuang2014}. More recently, the entanglement entropy of
general St\"{u}ckelberg models both in AdS soliton and AdS black
hole backgrounds with full backreaction was studied in Ref.
\cite{Peng2014}. However, the models mentioned above are all in the
frame of Maxwell electromagnetic theory. It is of interest to
investigate holographic entanglement entropy in the nonlinear
electromagnetic generalization. We \cite{yao2014} have studied the
holographic entanglement entropy in the insulator/superconductor
phase transition for the Born-Infeld electrodynamics, and found that
the entanglement entropy increases with the increase of the
Born-Infeld factor in the superconductor phase and the critical
width $\ell$ of confinement/deconfinement phase transition is
dependent of the Born-Infeld parameter. As a further study along
this line, in this paper we will extend the previous works to investigate the
entanglement entropy in metal/superconductor phase transition with Born-Infeld
electrodynamics. We find that the
the entanglement entropy decreases with the increase of
the Born-Infeld parameter $b$ for both operators in the metal phase.
However, In the superconductiong phase, the entanglement entropy of operator
$<\mathcal{O}_{-}>$ increases with the increase of Born-Infeld factor. Interestingly,
for the operator $<\mathcal{O}_{+}>$, the effect of the Born-Infeld
parameter on the entanglement entropy is non-monotonic and the value of the
threshold $b_{T}$ is related to the belt width of subsystem $A$.

The framework of this paper is as follows. In Sec. II, we will
derive the equations of motions and introduce the boundary conditions of
the holographic model. In Sec. III, we study
the phase transition with Born-Infeld electrodynamics in
AdS black hole spacetime. In Sec. IV, we calculate the
holographic entanglement entropy in AdS black hole gravity with
Born-Infeld electrodynamics. In Sec. V, we conclude our main
results of this paper.

\section{Equations of motion and boundary conditions}
The action for the gravity and Born-Infeld electromagnetic field coupling
with a charged scalar field is described by
\begin{eqnarray}\label{action}
S&=&\int d^{d}
x\sqrt{-g}\left[\frac{1}{16\pi
 G}(R-2\Lambda)\right]
+\int d^{d} x\sqrt{-g}\Bigg[\frac{1}{b^2}\left(1-\sqrt{1+\frac{b^2
F^{\mu\nu}F_{\mu\nu}}{2}}\right)\nonumber \\
&&
-|\nabla\Psi-i q A\Psi
|^2-m^2|\Psi|^2\Bigg],
\end{eqnarray}
where $g$ is the determinant of the metric, $\psi$ represents a scalar field with charge $q$ and mass $m$, $\Lambda=-(d-1)(d-2)/2L^2$ is the cosmological constant, $A$ is the gauge field, $F^{\mu\nu}$ is the strength of the Born-Infeld electrodynamic field $F=d A$,
and $b$ is the Born-Infeld coupling parameter. In the limit $b\rightarrow0$, the Born-Infeld field will reduce to the Maxwell field.

To study the holographic entanglement entropy in Born-Infeld electrodynamics, we will take the full backreaction into account. The metric for the planar black hole can be taken as
\begin{eqnarray}\label{BH metric}
ds^2=-f(r)e^{-\chi(r)}dt^{2}+\frac{dr^2}{f(r)}+r^{2}h_{ij}dx^{i}dx^{j}.
\end{eqnarray}
The Hawking temperature of this black hole is
\begin{eqnarray}\label{Hawking temperature}
T_{H}=\frac{f^{\prime}(r_{+})e^{-\chi(r_{+})/2}}{4\pi},
\end{eqnarray}
where $r_{+}$ is the black hole horizon.

We consider the electromagnetic field and the scalar field in the
forms
\begin{eqnarray}
A=\phi(r)dt,~~\psi=\psi(r).
\end{eqnarray}
Then, the equations of motion from the variation of the action with respect to the matter
and metric can be obtained as
\begin{eqnarray}\label{psi}
&& \psi^{\prime\prime}+\left(\frac{d-2}{r}-\frac{\chi^{\prime}}{2}+
\frac{f^\prime}{f}\right)\psi^\prime+\frac{1}{f}\left(\frac{q^{2}e^{\chi}\phi^2}{f}-m^2\right)
\psi=0,
\\ \label{phi}
&& \phi^{\prime\prime}+\left(\frac{d-2}{r}+\frac{\chi^{\prime}}{2}\right)\phi^\prime-
\frac{(d-2)b^2e^{\chi}}{r}\phi'^3-
\frac{2q^{2}\psi^{2}(1-b^2 e^{\chi}\phi'^2)^\frac{3}{2}}{f}\phi=0,
\\
&& \chi^{\prime}+\frac{2r}{d-2}\left(\psi^{\prime
2}+\frac{q^{2}e^{\chi}\phi^{2}\psi^{2}}{f^{2}}\right)=0,\label{chi}
\end{eqnarray}
\begin{eqnarray}
&& f^{\prime}-\left(\frac{(d-1)r}{L}-\frac{(d-3)f}{r}\right)+\frac{r}{d-2}
\left[m^{2}\psi^{2}
+f\left(\psi^{\prime
2}+\frac{q^{2}e^{\chi}\phi^{2}\psi^{2}}{f^{2}}\right)+
\frac{1-\sqrt{1-b^2 e^{\chi}\phi'^2}}{b^2\sqrt{1-b^2 e^{\chi}\phi'^2}}\right]=0,\nonumber \\
\end{eqnarray}
where a prime denotes the derivative with respect to $r$, and $16\pi G=1$ was used. For this system, the useful scaling symmetries are
\begin{eqnarray}
&& r\rightarrow \alpha r,\qquad (x,y,t)\rightarrow(x,y,t)/\alpha,\qquad\phi\rightarrow \alpha\phi,\qquad f\rightarrow\alpha^2f,\label{scaling:r}\\
&& L\rightarrow\alpha L,\qquad r\rightarrow \alpha r,\qquad t\rightarrow \alpha t,\qquad q\rightarrow\alpha^{-1} q, \label{scaling:L}\\
&& e^{\chi}\rightarrow\alpha^2e^{\chi},\qquad\phi\rightarrow \alpha^{-1}\phi,\qquad t\rightarrow t\alpha. \label{scaling3}
\end{eqnarray}
Using the scaling symmetries (\ref{scaling:r}), we can take $r_+=1$.
Further employing the symmetries (\ref{scaling:L}), we can let
$L=1$.

At the horizon $r_+$, the regularity condition  gives the boundary conditions
\begin{eqnarray}
\phi(r_+)=0,\qquad f(r_+)=0.
\end{eqnarray}
And at the asymptotic AdS boundary ($r\rightarrow\infty$), the
asymptotic behaviors of the solutions are
\begin{eqnarray}
\chi\rightarrow0\,,\hspace{0.5cm}
f\sim r^2\,,\hspace{0.5cm}
\phi\sim\mu-\frac{\rho}{r^{d-3}}\,,\hspace{0.5cm}
\psi\sim\frac{\psi_{-}}{r^{\Delta_{-}}}+\frac{\psi_{+}}{r^{\Delta_{+}}}\,,
\label{infinity}
\end{eqnarray}
where $\mu$ and $\rho$ are interpreted as the chemical potential and
charge density in the dual field theory, and the exponent $\Delta_\pm$ is defined by
$((d-1)\pm\sqrt{(d-1)^2+4m^{2}})/2$ for 4-dimensional spacetime. Notice that, provided $\Delta_{-}$ is larger than the unitarity
bound, both $\psi_{-}$ and $\psi_{+}$ can be normalizable. According to the AdS/CFT correspondence, they correspond to the vacuum expectation values $\psi_{-}=<\mathcal{O}_{-}>$, $\psi_{+}=<\mathcal{O}_{+}>$ of an operator $\mathcal{O}$ dual to the scalar field
\cite{HartnollPRL101,HartnollJHEP12}. In the following calculation, we impose boundary condition that either  $\psi_{-}$ or $\psi_{+}$ vanishes.

\section{phase transition with Born-Infeld electrodynamics}
In this section, we will concretely study the phase transition for Born-Infeld electrodynamics with full backreaction in
the four-dimensional AdS black hole spacetime. For the normal phase, the metric becomes the Reissner-Nordstr\"{o}m
AdS black hole as the Born-Infeld factor approaches to zero. Thus,
we have
\begin{equation}\label{RN AdS}
\chi=\psi=0, \qquad \phi=\rho\left(1-\frac{1}{r}\right),\qquad
f=r^2-\frac{1}{r}\left(1+\frac{\rho^2}{4}\right)+\frac{\rho^2}{4r^2}.
\end{equation}
However, if the Born-Infeld factor is not equal to zero, the
solution is the Born-Infeld AdS black hole.

For purpose of getting the
solutions in superconducting phase where $\psi(r)\neq0$, we can
introduce a new variable $z=r_{+}/r$. Then, the equations of motion
can be rewritten as
\begin{eqnarray}\label{psiz}
&&\psi^{\prime\prime}-\left(\frac{\chi^{\prime}}{2}-
\frac{f^\prime}{f}\right)\psi^\prime
-\frac{1}{z^{4}f}\left(m^2-\frac{e^{\chi}q^2\phi^2}{f}\right)\psi=0,
\\
&&\phi^{\prime\prime}+\frac{1}{2}\chi^{\prime}\phi^\prime+2z^3b^2e^{\chi}\phi'^3-\frac{2q^{2}\psi^{2}(1-b^2
e^{\chi}z^4\phi'^2)^\frac{3}{2}}{z^{4}f}\phi=0,
\\
&&\chi^{\prime}-z\psi^{\prime
2}-\frac{e^{\chi}q^2\phi^{2}\psi^{2}}{z^{3}f^{2}}=0,\label{chiz}
\\ \label{frz}
&&f^{\prime}-\frac{f}{z}+\frac{3r^{2}_{+}}{z^{3}}-\frac{1}{2z^{3}}
\left[m^{2}\psi^{2}+f\left(z^{4}\psi^{\prime
2}+\frac{1}{f^{2}}e^{\chi}q^2\phi^{2}\psi^{2}\right)+
\frac{1-\sqrt{1-b^2z^4e^{\chi}\phi'^2}}{b^2\sqrt{1-b^2z^4 e^{\chi}\phi'^2}})\right]=0,\nonumber \\
\end{eqnarray}
where the prime now denotes the derivative with respect to  $z$.
Using the shooting method, we can solve the equations of motion
numerically and then discuss the effects of the Born-Infeld
parameter $b$ on the condensation of the scalar operators. In this
paper, we set $m^2=-2$ and $q=1$. Since there are scaling symmetries
described by Eq. (\ref{scaling:r}) for the equations of motion, the
following quantities can be rescaled as
\begin{equation}\label{scaling Q}
\mu\rightarrow\alpha\mu,\qquad
\rho\rightarrow\alpha^2\rho, \qquad
<\mathcal{O}_{-}>\rightarrow\alpha <\mathcal{O}_{-}>,\qquad
<\mathcal{O}_{+}>\rightarrow\alpha^{2}<\mathcal{O}_{+}>.
\end{equation}

\begin{figure}[ht]
\center{
\includegraphics[scale=1.0]{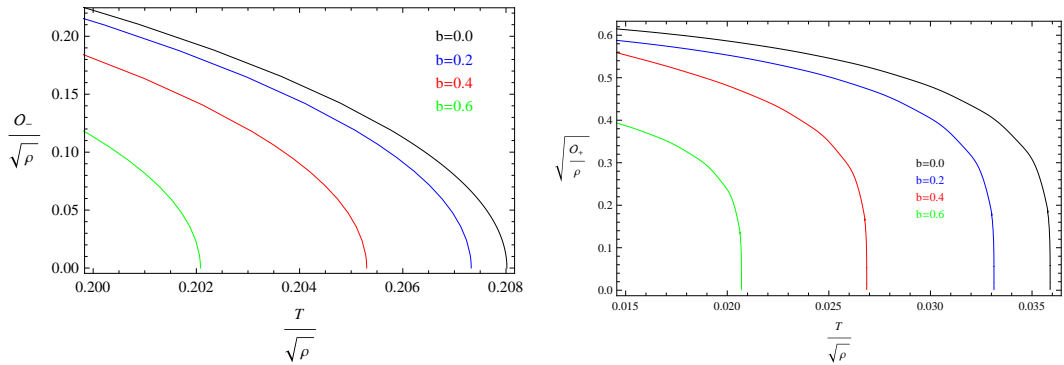}
\caption{ (Color online.) The operators $O_-$ (left plot) and $O_+$
(right plot) versus temperature after condensation with different
values of Born-Infeld parameter. The four lines from top to bottom
correspond to $b=0$ (black), $b=0.2$ (blue), $b=0.4$ (red), and
$b=0.6$ (green) respectively.} \label{condesation} }
\end{figure}

In Fig.~\ref{condesation}, we plot the behaviors of condensate with
the changes of the temperature and the Born-Infeld parameter in the
dimensionless quantities $<\mathcal{O}_{-}>/\sqrt{\rho},$
$\sqrt{<\mathcal{O}_{+}>/\rho}$ and $T/\sqrt{\rho}$.
From the left plot of Fig.~\ref{condesation}, we can see that if the
temperature $T>T_c$, there is no condensation and this can be
thought of as the metal phase. However, when the temperature
decreases to be lower than the critical value $T_c$, the
condensation of the operator $<\mathcal{O}_{-}>$ emerges and this
corresponds to a superconducting phase. It should be noted that the
value of the critical temperature $T_c$ becomes smaller as the
Born-Infeld factor $b$ increases, which means that the Born-Infeld
correction to the usual Maxwell field makes the scalar hair harder
to form in the full-backreaction model. For the operator
$<\mathcal{O}_{+}>$ (right plot), we also find that the critical
temperature $T_c$ decreases with the increase of the parameter $b$.
Our result is the same as the result in Ref. \cite{HartnollJHEP12}
as the factor $b$ approaches to zero.

\section{Holographic entanglement entropy of the holographic model}
After obtaining the solutions to the metal phase and superconducting phase
for the Born-Infeld electrodynamics in the
the AdS black hole geometry with full backreaction,
we are now ready to study the behavior of holographic entanglement entropy in the
Born-Infeld electrodynamics. The entanglement entropy in conformal
field theories can be calculated from the area of minimal surface
in AdS spaces \cite{Ryu:2006bv, Ryu:2006ef}, and its formula is
given by the ``area law"
\begin{equation}\label{law}
S_A=\frac{\rm Area(\gamma_\mathcal{A})}{4G_N},
\end{equation}
where $G_N$ is the Newton constant in the Einstein gravity on the AdS space, $S_A$ is the entanglement
entropy for the subsystem $A$ which can be chosen arbitrarily, $\gamma_A$ is the minimal
area surface in the bulk which ends on the boundary of $A$.

We consider the entanglement entropy for a straight geometry which is described by
$-\frac{l}{2}\leq x \leq \frac{l}{2}\ $ and $-\frac{R}{2}<y<\frac{R}{2}~(R\rightarrow\infty)$,
where $\ell$  is defined as the size of region $A$.
The holographic surface $\gamma_A$ starts from $r=\frac{1}{\epsilon}$ at $x=\frac{\ell}{2}$, extends into the bulk until it reaches $r=r_*$, then returns back to the AdS boundary $r=\frac{1}{\epsilon}$ at $x=-\frac{\ell}{2}$.
Thus, the induced metric on $\gamma_A$ can be obtained as follows
\begin{equation}
ds^2 =\left[\frac{1}{f(r)}+r^2\left (\frac{dx}{dr}\right )^2\right
]dr^2+r^2 dy^2.
\end{equation}
By using the proposal given by Eq. (\ref{law}), the entanglement entropy in the strip geometry is
\begin{eqnarray}\label{EEntropyBH}
S_A=\frac{R}{2G_4}\int^{z_{*}}_{\varepsilon}dz\frac{z_{*}^{2}}{z^{2}}\frac{1}{\sqrt{(z^{4}_{*}-z^{4})z^{2}f(z)}}
=\frac{R}{2G_4}\left(\frac{1}{\varepsilon}+s\right),
\end{eqnarray}
with
\begin{eqnarray}\label{Length}
\frac{l}{2}=\int^{z_{*}}_{\varepsilon}dz\frac{z^{2}}{\sqrt{(z^{4}_{*}-z^{4})z^{2}f(z)}},
\end{eqnarray}
where $z_{*}$ satisfies the condition $\frac{dz}{dx}|_{z_{*}}=0$
with $z=\frac{1}{r}$.   The term $1/\epsilon$ in Eq. (\ref{EEntropyBH}) is divergent, while the term $s$ is a
finite term which is physically important. Under the scaling symmetries of Eq. (\ref{scaling:r}), we can rescale the $\ell$ and $s$ as
\begin{equation}\label{scaling s}
\ell\rightarrow\alpha^{-1}\ell,\quad s\rightarrow\alpha s.
\end{equation}
Therefore,  in the following calculation we can use these dimensionless quantities
\begin{equation}\label{Ls}
\ell\sqrt{\rho}, \qquad s/\sqrt{\rho}.
\end{equation}
We now show the behavior of the holographic entanglement entropy of
the operator $<O_{-}>$ or $<O_{+}>$ with respect to the temperature
$T$, the Born-Infeld factor $b$ and the belt width $\ell$,
respectively.

\subsection{Holographic entanglement entropy of the operator $<O_{-}>$}

\begin{figure}[ht]
\center{
\includegraphics[scale=0.6]{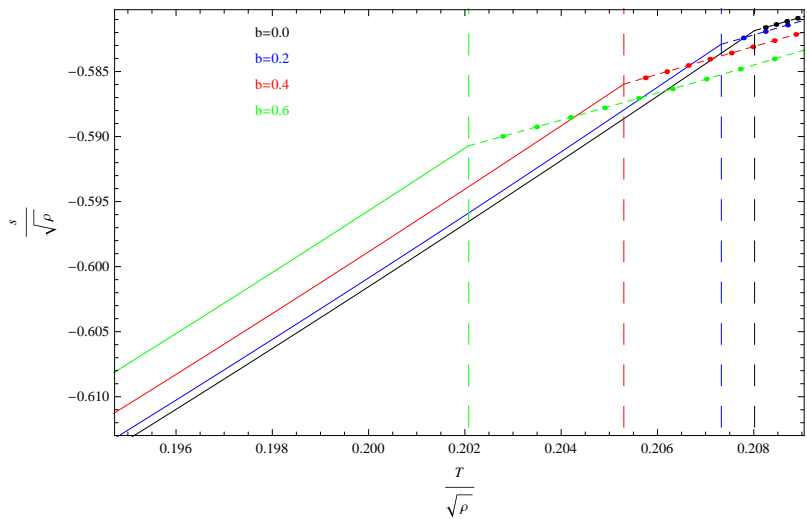}
\caption{\label{gfst}(Color online.) The entanglement entropy of the
operator $<O_{-}>$ as a function of the temperature $T$ and the
Born-Infeld factor $b$ for  $\ell\sqrt{\rho}=1$. The four dot-dashed
lines from top to bottom correspond to $b=0$ (black), $b=0.2$
(blue), $b=0.4$ (red), $b=0.6$ (green), but the four solid lines
from bottom to top are for $b=0$ (black), $b=0.2$ (blue), $b=0.4$
(red), and $b=0.6$ (green) respectively.}}
\end{figure}

The behavior of the entanglement entropy of the operator $<O_{-}>$
is shown in Fig. \ref{gfst} in which the dot-dashed lines describe
the normal phases and the solid ones show the superconducting
phases. It can be seen from the figure that the entanglement entropy
presents a discontinuous change at a critical temperatures $T_c$
denoted by vertical dashed lines for different strengths of the
Born-Infeld parameter $b$. The discontinuous change of the
entanglement entropy indicates the phase transition point from the
normal state to the superconducting state and the value of the
$T_c$ becomes smaller with the increase of the Born-Infeld factor
$b$. Which indicates that the holographic
entanglement entropy is a good probe to study the properties of the
phase transition.
The figure also shows that the entanglement entropy
decreases as the Born-Infeld parameter increases for the normal
phase, but increases as the Born-Infeld parameter increases for the
superconducting state.  Moreover, the
entanglement entropy in superconducting phase is less than the one
in the normal case and drops as the temperature decreases, and this
property holds for different values of the parameter $b$. This
behavior of the entanglement entropy is due
to the fact that the metal phase can be thought of as the one filled with free charge carriers,
such as electrons. The condensate turns on at the critical temperature and the free charge
carriers are continuously condensed to Cooper pairs as temperature decreases. Therefore,
the formation of Cooper pairs make the degrees of
freedom decrease in the superconducting phase.

\begin{figure}[ht]
\center{
\includegraphics[scale=0.9]{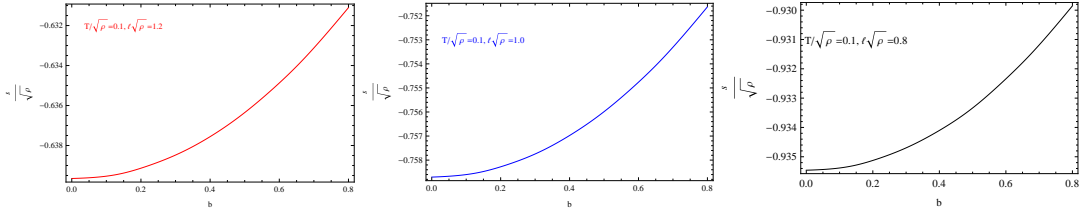}
\caption{\label{fsb}(Color online.) The entanglement entropy of the
operator $<O_{-}>$ as a function of the Born-Infeld factor $b$ for
different widths $\ell$ as $T/\sqrt{\rho}=0.10$. The left plot (red)
is for $\ell\sqrt{\rho}=1.2$, the middle one (blue) for
$\ell\sqrt{\rho}=1.0$, and the right one (black) for
$\ell\sqrt{\rho}=0.8$.}}
\end{figure}

The entanglement entropy as a function of the parameter $b$ for
different $\ell$ as $T/\sqrt{\rho}=0.10$ is shown in Fig. \ref{fsb}. For fixed belt width $\ell$, the
entanglement entropy becomes smaller as the Born-Infeld factor $b$
decreases. This is because that the condensation becomes stronger
with higher condensation gap for smaller parameter $b$ at low
temperature so that the number of Cooper pairs is increased, which
results in less degree of freedom. On the other hand, for a given
Born-Infeld factor $b$, with the decrease of belt width $\ell$ the
entanglement entropy also decreases.

\subsection{Holographic entanglement entropy of the operator $<O_{+}>$}

\begin{figure}[ht]
\center{
\includegraphics[scale=1.1]{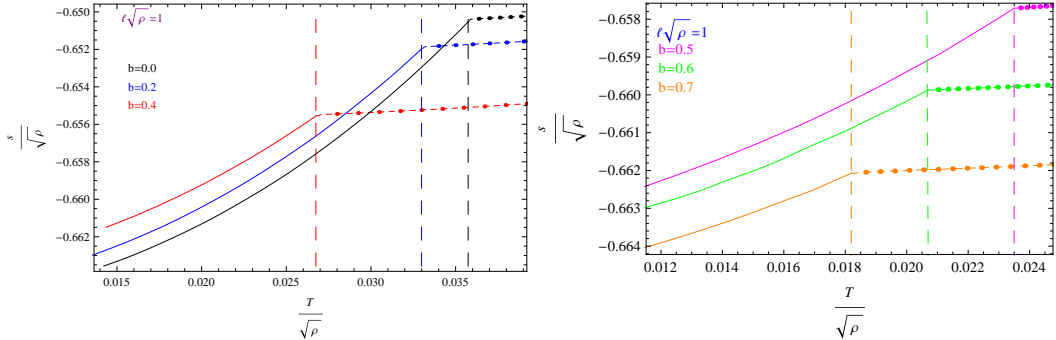}
\caption{\label{gzst}(Color online.) The entanglement entropy of the
operator $<O_{+}>$ as a function of the temperature $T$ and the
Born-Infeld factor $b$ with $\ell\sqrt{\rho}=1$. The vertical dashed
lines represent the critical temperature of the phase transition for
the different values of the Born-Infeld factor. The dot-dashed lines
are from the normal phase and the solid ones are from the
superconducting cases. In the left plot, the three lines from bottom
to top correspond to $b=0$ (black), $b=0.2$ (blue) and $b=0.4$
(red), but in the right one are for $b=0.7$ (orange), $b=0.6$
(green), $b=0.5$ (magenta), respectively.}}
\end{figure}

The behavior of the entanglement entropy of the operator $<O_{+}>$
as a function of temperature $T$ and the Born-Infeld factor $b$ is
described by Fig. \ref{gzst}.  It is shown that the critical
temperature $T_c$ of the phase transition decreases as the
Born-Infeld factor $b$ increases. That is to say, the stronger
Born-Infeld electrodynamics correction makes the scalar condensation
harder to form. We also find that the change of the entanglement
entropy is discontinuous at $T_c$ and the entanglement entropy in
the hair phase is less than the one in the normal phase.
Interestingly, in the superconducting phase, the dependence of the
entanglement entropy on the Born-Infeld factor $b$ is non-monotonic.
In the left plot of Fig. \ref{gzst}, we find that the entanglement
entropy increases with the increase of the factor $b$ when
$b<b_{T}$. However, in the right plot of Fig. \ref{gzst}, we can see
that the entanglement entropy decreases with the increase of the
factor $b$ when $b>b_{T}$ .

\begin{figure}[ht]
\includegraphics[scale=0.9]{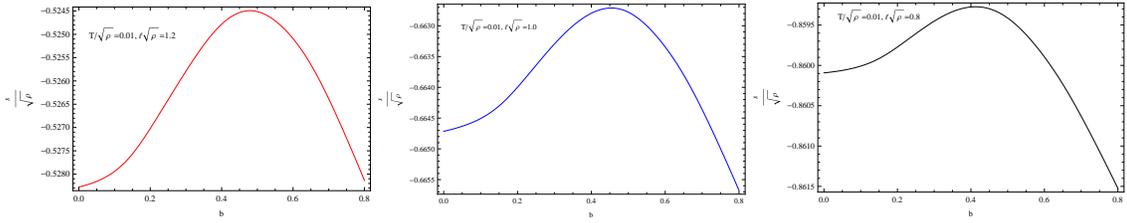}
\caption{\label{gzsb}(Color online.) The entanglement entropy of the
operator $<O_{+}>$ as a function of the Born-Infeld factor $b$ for
different widths $\ell$ with $T/\sqrt{\rho}=0.010$. The left plot
(red) is for $\ell\sqrt{\rho}=1.2$, the middle one (blue) for
$\ell\sqrt{\rho}=1.0$, and the right one (black) for
$\ell\sqrt{\rho}=0.8$.}
\end{figure}

To further illustrate the effect of the Born-Infeld factor $b$ on
the entanglement entropy of the operator $<O_{+}>$ in the
superconducting phase, we plot the entanglement entropy of the
operator $<O_{+}>$ as a function of Born-Infeld factor $b$ for
different widths $\ell$ with $T/\sqrt{\rho}=0.010$ in Fig.
\ref{gzsb}. Obviously, with the increase of the factor $b$, the
entanglement entropy first rises and arrives at its maximum as
$b=b_{T}$, then decreases monotonously. This process implies that
there is some kind of the significant reorganization of the degrees
of freedom. And the threshold $b_{T}$ becomes smaller as the width
of the subsystem $\ell$ decreases. For the fixed Born-Infeld factor
$b$, the entanglement entropy decreases with the decrease of belt
width $\ell$.

\section{Summary}
We studied the behaviors of the holographic entanglement entropy in the metal/superconductor phase transition for the Born-Infeld electrodynamics with full backreaction. By calculating the entanglement entropy of the system, we noted that the critical temperature of the condensation for the operators $<\mathcal{O}_{-}>$ and $<\mathcal{O}_{+}>$ becomes smaller with the increase of the Born-Infeld parameter $b$, which implies that the Born-Infeld factor makes the scalar condensation harder to form. Both for the operators $<\mathcal{O}_{-}>$ and $<\mathcal{O}_{+}>$, we found that the entanglement entropy in the superconducting phase is less than the one in the normal phase, and drops as the temperature decreases for the fixed parameter $b$ and belt width
$\ell$. This is due to the fact that the formation of Cooper pairs makes the degrees of freedom decrease in the hair phase. For a given temperature, we observed that the entanglement entropy of the operator $<\mathcal{O}_{-}>$ in the metal (or superconducting) phase decreases (or increases) with the increase of the Born-Infeld factor $b$ for the fixed $\ell$.  Interestingly, the influence of the Born-Infeld factor $b$ on the entanglement entropy of the operator $<\mathcal{O}_{+}>$  is nontrivial. The entanglement entropy of the operator $<\mathcal{O}_{+}>$ in the metal phase decreases with the increase of the Born-Infeld parameter. However, with the increase of the Born-Infeld parameter $b$ for the fixed belt width $\ell$ of subsystem $A$, the entanglement entropy of the operator $<\mathcal{O}_{+}>$ in the superconducting phase first increases and reaches the maximum at some threshold  $b_{T}$, then decreases monotonously. The threshold $b_{T}$ becomes smaller as the width of the subsystem $A$ decreases. This process implies that there is some kind of the significant reorganization of the degrees of freedom which should be further studied in the future.

\begin{acknowledgments}
This work was supported by the National Natural Science Foundation
of China under Grant No. 11175065; the SRFDP under Grant No.
20114306110003; Hunan Provincial Innovation Foundation For
Postgraduate under Grant No. CX2013A009; and Construct Program of
the National Key Discipline.
\end{acknowledgments}

\end{document}